# Two-dimensional $PS_2$: a promising anode material for sodium-ion batteries and a potential superconductor

Dawei Zhou[a], Xin Tang[b], Chunying Pu[a,*]

[a] College of Physics and Electronic Engineering, Nanyang Normal University, Nanyang 473061, China
[b] College of Material Science and Engineering, Guilin University of Technology, Guilin 541004, China

**Abstract**

Two dimensional materials as electrodes have shown unique advantages such as the infinite planar lengths, broad electrochemical window, and much exposed active sites. In this work, by means of density functional theory computations, we demonstrate that two dimensional $PS_2$ with the 1T-Type structure as many two dimensional disulfides is a promising anode material for sodium ion batteries application. Different from many two dimensional disulfides (e.g. $MoS_2$, $TiS_2$, $CrS_2$) compounds that are semiconducting, $PS_2$ monolayer exhibits metallic character with considerable electronic states at the Fermi level, which can provide good electrical conductivity during the battery cycle and also suggest the potential superconductivity. Remarkably, $PS_2$ monolayer has a considerably high theoretical capacity of 1692 mAh/g, a rather small sodium diffusion barrier of 0.17 eV, and a low average open circuit voltage of 0.18 V. These results suggest that $PS_2$ monolayer can be utilized as a promising anode material for the application in sodium ion batteries with high power density and fast charge/discharge rates. In addition, estimated by first principle calculations, $PS_2$ is found to be an intrinsic phonon-mediated superconductor with a relatively high critical superconducting temperature of about ~10 K.

†Correspondence should be addressed to: Chunying Pu (puchunying@126.com.)

# 1. Introduction

As the clean energy storage technologies, Li-ion batteries (LIBs) exhibit many advantages such as high energy density, great power density, structural flexibility and stability, environmentally friendly and so on.[1-5] So LIBs have achieved great commercial success especially in the portable device market in past ten years.[6-7] However, the lack of lithium resources in the earth and the limited storage capacity impeders its further applications in large-scale.[8] In order to meet the demand for the next generation metal-ion batteries, extensive efforts have been made to develop new metal ion batteries.[9-17] Among various anode material candidates for metal ion batteries, two dimensional sodium-ion batteries (SIBs) have been paid more attentions.[12-17] Firstly, sodium is more abundant and more safety than lithium. Secondly, due to a large surface-area-to-volume ratio, two dimensional (2D) materials have a larger contact area between the electrolyte and electrode, which usually favor the improvement of energy density. So 2D SIBs are believed to be promising candidates for the next generation anode materials in metal ion batteries. Unfortunately, the high-performance anode materials in well developed LIBs generally are not applicable to SIBs because of the larger atomic radius of Na compared with Li. For example, by using Na instead of Li, the storage capacity of graphite falls from 372 mAh/g to just 35 mAh/g.[18,19] Therefore, design and preparation of 2D SIBs anode materials with desirable properties is indispensable.

Theoretical calculations based first-principle method play an important role in understanding the charge/discharge mechanisms and electronic properties of electrode materials. To date, a tremendous number of 2D materials, including graphene systems,[20,21] phosphorene,[22-23] MXene,[24-25] transition-metal dichalcogenides and nitrides,[26-30] have been predicted to exhibit excellent performance in SIBs. For example, phosphorene as anode in SIBs achieves the theoretical capacity of 865 mAh/g,[22] and borophene gain a maximum theoretical capacity of 1984 mAh/g.[31] Other 2D materials which are studied by DFT calculations as potential anode materials for SIBs are defective graphene (1450 mAh/g),[32] borocarbonitride based anode (810 mAh/g),[33] silicene, germanene and stanene (954, 369, and 226 mAh/g, respectively),[34,35] B-doped graphene (762 mAh/g),[36] double layer graphene-phosphorene hybrid (DG/P) (372 mAh/g).[37] The ion diffusion of most these 2D materials is roughly between 0.1 to 0.6 eV. So searching 2D materials with good performance for applications in SIBs still face great challenge such as volume expansion problem, capacity restrict, diffusion barrier issue. Designing more 2D materials suitable for SIBs are still necessary.

In this work, through first-principle calculation we study the $PS_2$ monolayer as two-dimensional and investigate their electronic properties for superior anode materials of NIBs. The stability of $PS_2$ monolayer is confirmed through the phonon spectra, ab initio molecular dynamics (AIMD) simulations, and in-plane stiffness constants. Our calculations to the electronic properties show that the $PS_2$ monolayer is metallic, which is advantageous for the applications in Na-ion batteries. The storage capacity of the $PS_2$ monolayer as Na-ion material is 1692 mAh/g, corresponding the ion diffusion barrier of 0.17 eV and the open current voltage of 0.18 V. Furthermore, we find that the $PS_2$ monolayer is also exhibits superconducting behavior with the ~10 K superconductivity transition temperature by the calculations of electron-phonon coupling.

## 2. Computational Methods

The first-principles calculation are done with the projector augmented wave (PAW) method[38,39] as implemented in the Vienna ab initio simulation package

†Correspondence should be addressed to: Chunying Pu (puchunying@126.com.)

(VASP).[40,41] The electron exchange-correlation energy was treated within the generalized gradient approximation (GGA), using the functional of Perdew, Burke, and Ernzerhof (PBE).[42] The energy cutoff of the plane wave was set to 380 eV and the Brillouin zone was sampled with a 12×12×1 Monkhorst-Pack k-point grid. All the atomic positions are fully optimized with the convergence of $10^{-5}$ eV and $10^{-3}$ eV/Å for energy and force, respectively. A large vacuum space of 35 Å in the perpendicular direction of the sheet is used to avoid the interactions between periodic images. Phonon dispersion calculations were based on a supercell approach as used in the Phonopy code.[43] We take the 3×3 supercell for calculating the phonon spectra of the $PS_2$ monolayer. In order to determine the dynamical stability of the $PS_2$ monolayer, the thermal stability was analyzed by ab initio molecular dynamics (AIMD) simulations using the canonical ensemble (NVT) with a 3×3×1 supercell. In the calculation of sodium ions diffusion, we used nudged elastic band (NEB) method to get the ion diffusion/barrier.[44]

The electron-phonon coupling (EPC) constants λ and the superconducting transition temperature were calculated in the framework of density functional perturbation theory as implemented in the QUANTUM ESPRESSO codes.[45] The ultrasoft pseudopotentials[46] and the GGA exchange-correlation potentials[42] were used to model the electron-ion interactions. After the full convergence test, the energy cutoff of the plane wave was set to be 60 Ry, and the Brillouin-zone mesh of 24 × 24 × 1 points for the self-consistent electron density calculation was used. The EPC coefficients were further calculated with a 12× 12 × 1 mesh of q-points.

## 3. Results and discussion

Monolayer $PS_2$ is one of the $AB_2$ structure[47] and we found that it is energetically more stable in its 1T phase than its 1H phase. As shown in Figure 1a, the structure of

PS$_2$ consist of three atomic sub-planes similar to 1T-MoS$_2$. The sub-plane of P atoms is sandwiched between the two sub-planes of sulfur atoms. The optimized lattice constants are $a=b=3.288$ Å with a layer thickness of 2.72 Å and the distance of P-S is 2.334 Å. To evaluate the chemical bonding, we computed the charge difference density, which is defined as the total electronic density of the PS$_2$ monolayer minus the electron density of isolated P and S atoms at their respective positions. It is obviously seen that the non-polar covalent bonding character is evidenced by the presence of the electron density between P-S (Fig.1b). According to the Bader charge population analysis, the P atom and S atom in PS$_2$ monolayer possess a charge of 0.88 and -0.44 |$e$|, respectively.

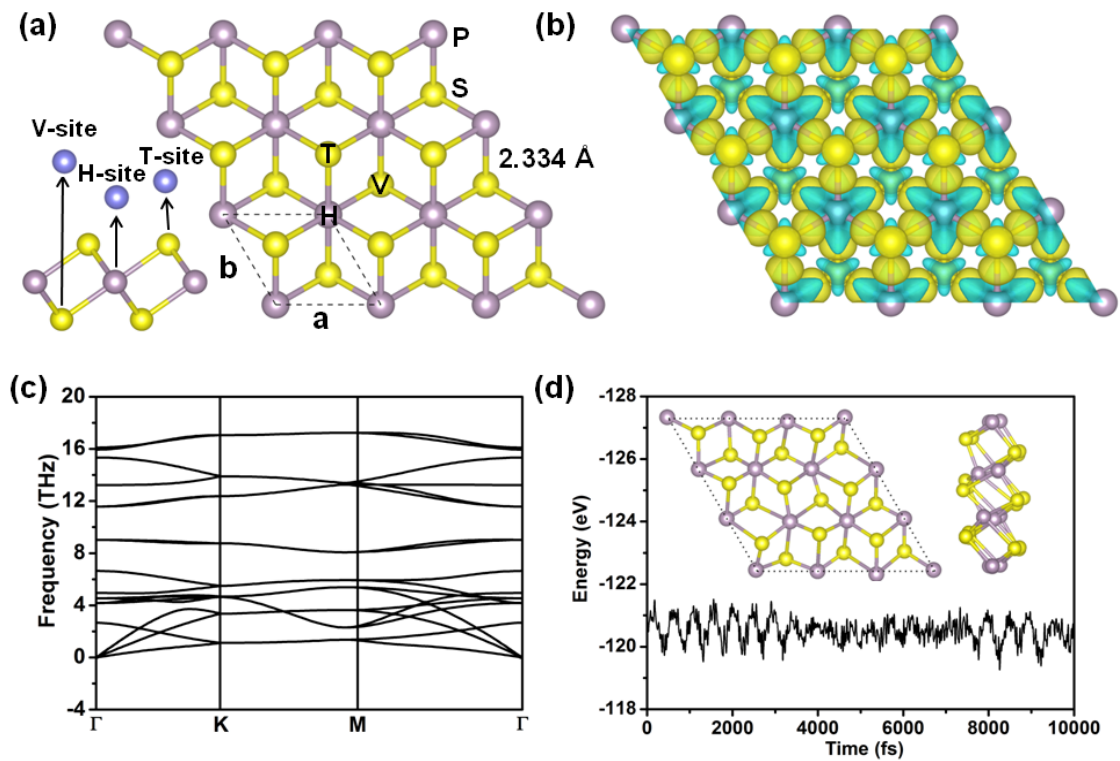

Figure 1. (a)The structure of two dimensional PS$_2$ monolayer. P and S atoms are represented by gray and yellow spheres, respectively. (b) Difference charge density of PS$_2$. The gold color (i.e., 0.005 eÅ$^{-3}$) in the plot indicates an electron density increase in the electron density after bonding, and the cyan color (i.e., 0.005 eÅ$^{-3}$) indicates an loss. (c) Phonon dispersion curves of PS$_2$ monolayer. (d) Fluctuation of total potential

†Correspondence should be addressed to: Chunying Pu (puchunying@126.com.)

energy of the PtS$_2$ during the AIMD simulation at 600 K. The inset is the structure of PS$_2$ monolayer at the end of the AIMD simulation.

To explore the energetic stability of the PS$_2$ monolayer, the binding energy is calculated, which is defined by the following formula:

$$E_b = (E_P + 2E_S - E_{PS_2})/3 \qquad (1)$$

where $E_P(E_S)$ and $E_{PS_2}$ represents the total energies of a single P(S) atom and PS$_2$ monolayer, respectively. The binding energy of the PS$_2$ monolayer is 4.54 eV/atom, higher than that of silicene and germanene (3.98 and 3.26 eV/atom, respectively),[48] suggesting that the P-S bond in PS$_2$ monolayer is robust. The dynamical stability of PS$_2$ monolayer was tested by calculating the phonon dispersion curves along the high-symmetry paths. As shown in Fig.1c, all vibrational modes are found to be real, confirming that PS$_2$ monolayer is dynamically stable. The highest frequency of the optical mode is up to 17.24 THz (~575cm$^{-1}$), which can comparable to those of black phosphorene (~450 cm$^{-1}$),[49] and MoS$_2$ (~500 cm$^{-1}$),[50] indicating the strong bonding characteristic in the PS$_2$ monolayer. To investigate the thermal stability of the PS$_2$ under ambient conditions, we performed AIMD simulations in NVT, running for 10 ps at 600 K with a time step of 1 fs. The fluctuation of the total potential energy with simulation time is plotted in Figure 1d, which shows that the average value of the total potential energy remains nearly constant during the entire simulation. The structure of PS$_2$ at the end of the simulation is also plotted in Figure 1d, revealing that the structure does not experience serious structure disruptions, which confirms that PS$_2$ monolayer possesses good thermal stability and can maintain structural stability at temperature of 600 K.

We further examine the mechanical stability of PS$_2$ monolayer by calculating its linear elastic constants. The elastic constants of PS$_2$ are $C_{11}=C_{22}=78$ Nm$^{-1}$, $C_{12}=45$

Nm$^{-1}$, and $C_{66}$=16.5 Nm$^{-1}$, respectively, which meet the necessary mechanical equilibrium conditions[51] for mechanical stability: $C_{11}C_{22} - C_{12}^2 > 0$ and $C_{11}$, $C_{22}$, $C_{66}$ > 0, confirming the mechanical stability of PS$_2$. We also calculate the in-plane Young's modulus (Y) of PS$_2$, which is defined as: $Y = (C_{11} - C_{12}^2)/C_{22}$. The Young's modulus of PS$_2$ is about 52 N/m. It is worth noting that this value is lower than that of silicene (62 N/m) and TiS$_2$ (74 N/m). [52,53] Therefore, the structure of PS$_2$ has better mechanical flexibility, which is propitious to the manufacture of flexible battery materials.

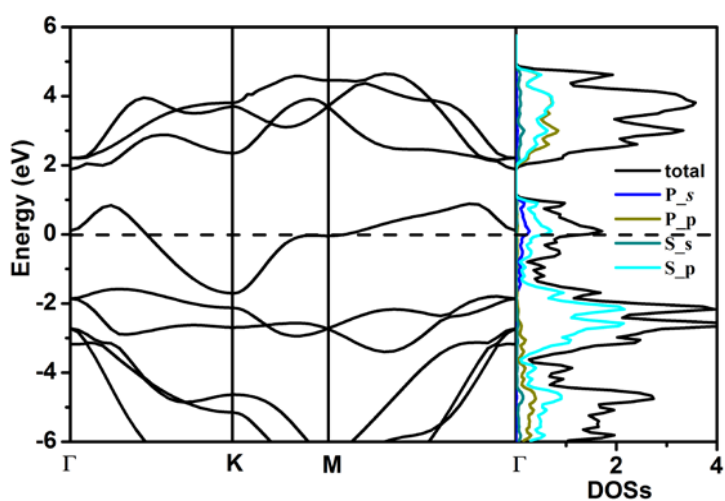

Fig.2 Electronic band structure calculated and projected density of states (DOSs) calculated by using the PBE functional.

The good electrical conductivity is essentially required for the excellent electrochemical performance of electrode materials and superconductivity. However, most of 2D disulfides(e.g. MoS$_2$, TiS$_2$, CrS$_2$)[53,54] are semiconductor with poor conductivity, which limits the electrochemical reactions and the generation of superconductivity. To better understand the nature of the electronic properties of PS$_2$, we calculate its electronic band structures and projected density of states. As shown in Figure 2, PS$_2$ exhibits metallic character with severe energy levels crossing the Fermi level. From the projected DOS analysis, the metallic states at the Fermi level are mainly contributed by S-2p states. Furthermore, there is also the contribution of P-2s orbitals at the Fermi level, indicating the formation of covalent bonds between P-S, which is consistent with the charge difference density analysis give above. Therefore, the metallic $PS_2$ monolayer is not only a highly desirable anode materials for SIBs but also imply a potential superconductor.

†Correspondence should be addressed to: Chunying Pu (puchunying@126.com.)

To estimate the superconducting transition temperature $T_C$, we use the Allen-Dynes formula [55]

$$T_C = \frac{\omega_{log}}{1.20} \exp(-\frac{1.04(1+\lambda)}{\lambda-\mu^*(1+0.62\lambda)}) \tag{2}$$

where $\omega_{log}, \lambda, \mu^*$ are the logarithmic average of the phonon energy, electron-phonon coupling constant and the electron-electron Coulomb repulsion parameter, respectively. The first two quantities were calculated as:

$$\omega_{log} = \exp(\frac{2}{\lambda}\int_0^\infty \alpha^2 F(\omega)\log\omega \frac{d\omega}{\omega}) \tag{3}$$

$$\lambda = 2\int_0^\infty \frac{\alpha^2 F(\omega)}{\omega} d\omega. \tag{4}$$

The electron-phonon spectral function $\alpha^2 F(\omega)$ appear in the expression, which is the central quantity in the superconductivity theory can be calculated in terms of the phonon linewidth as

$$\alpha^2 F(\omega) = \frac{1}{2\pi N(\varepsilon_F)}\sum_{qv}\delta(\omega-\omega_{qv})\frac{\gamma_{qv}}{\hbar\omega_{qv}} \tag{5}$$

where $N(\varepsilon_F)$, $\gamma_{qv}$ are the DOS at the Fermi level, and the linewidth of phonon mode v at the wave vector q. As shown in Figure 3, the Eliashberg spectral function $\alpha^2 F(\omega)$ and the electron-phonon coupling constant $\lambda(\omega)$ as functions of the phonon energy ω as well as the phonon dispersion and projected phonon density of states are plotted. We found λ=0.76, $\omega_{log}$=238 meV, and Tc=10.0 K (setting µ∗ to 0.1) for 2D $PS_2$. Since the modes having the same symmetry can be mixed with each other, the phonon eigenvectors have a strongly mixed character of P and S atoms. As clearly seen from Figure 3c, the contribution to the EPC constant λ come from the energy phonons with energies from 0 to 280 cm$^{-1}$. Due to the strong P-S interaction, we expect the out-of-plane vibrational modes sulfur coupled with the π* and interlayer states to produce superconductivity in PS$_2$ monolayer. The T$_c$ is about 10.0 K for PS$_2$, which is higher than lithium and calcium doped graphene (8.1 and 1.4 K, respectively)[56, 57] and compared to Li-intercalated bilayer MoS$_2$ superconductors (10.2 K).[58], but lower than intimate boron sheets (~12–21 K)[59,60] and B$_2$C (14.3–19.2 K).[61] The comparatively strong EP coupling λ and higher DOS at the Fermi level in the PS$_2$ monolayer are both favorable for the generation of superconductivity.

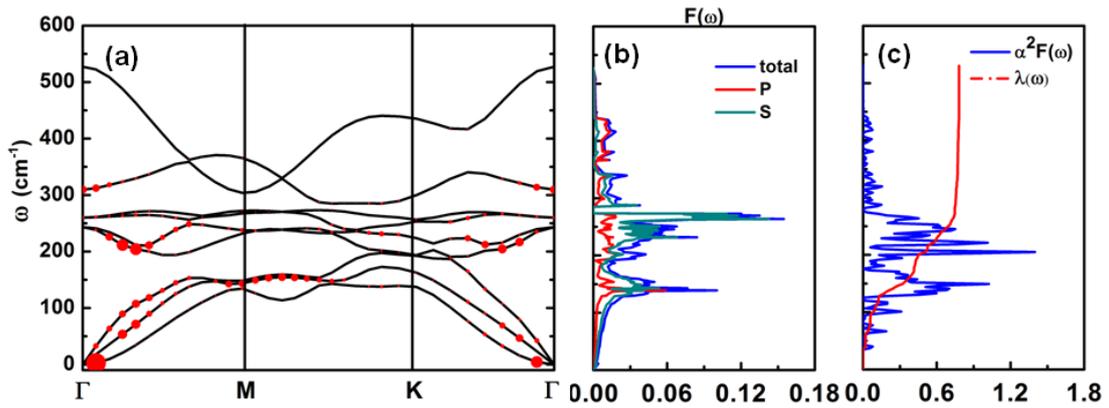

Figure 3 (a) Calculated phonon dispersion, (b) Phonon density of states, and (c) Eliashberg function α²F(ω) with interaged EP coupling constant λ(ω) for PS$_2$ monolayer.

Since the PS$_2$ monolayer is expected to be a potential anode materials in SIBs in view of its inherent metallicity, we firstly investigated the adsorption of a single Na atom on the surface of PS$_2$ monolayer by constructing a 2×2×1 supercell associated with the chemical stoichiometry of NaP$_4$S$_8$. The adsorption energy of Na is defined as:

$$E_{ad} = E_{PS_2Na} - E_{PS_2} - \mu_{Na} \quad (6)$$

†Correspondence should be addressed to: Chunying Pu (puchunying@126.com.)

where $E_{PS_2Na}$ and $E_{PS_2}$ are the total energies of the Na adsorbed $PS_2$ monolayer and pristine $PS_2$ monolayer, respectively, $\mu_{Na}$ is the chemical potential of Na and is taken as the cohesive energy of bulk Na. The negative value of adsorption energy means that the Na atom prefers to be adsorbed on the monolayer instead of forming a bulk metal and more favorable interaction between $PS_2$ monolayer and Na. Considering the lattice symmetry of $PS_2$ monolayer, three possible adsorption sites are considered, as shown in Figure 1a. After our geometrical optimization, three sites are remained and their adsorption energies are -0.89, -0.80, and -0.44 eV for V-, H-, and T-sites, respectively. The adsorption energies of three sites are negative, implying that Na atom prefers to be adsorbed on the host materials instead of forming a cluster.

To further understand the adsorption of Na atom, we make the Bader charge analysis and the Na atoms possess a charge of 0.82 and 0.83 |e| at V- and H-sites, respectively, which means that the charge transfer from Na atom to adjacent sulfur atoms. The existence of charge transference by Na atom reveals that the adsorption is chemical and can be regarded as redox reaction during the battery operation. The existence of chemical adsorption can be confirmed by the charge density difference (Figure 4c and Figure 4d), which is defined by $\Delta\rho = \rho(NaPS_2) - \rho(Na) - \rho(PS_2)$. The density of states of the $PS_2$ after adsorption of Na atom with V- and H-sites are also calculated, as shown in Figure 4a and 4b, the results show that the system still keeping metallic character, which is benefit for making electrode materials from the $PS_2$ monolayer.

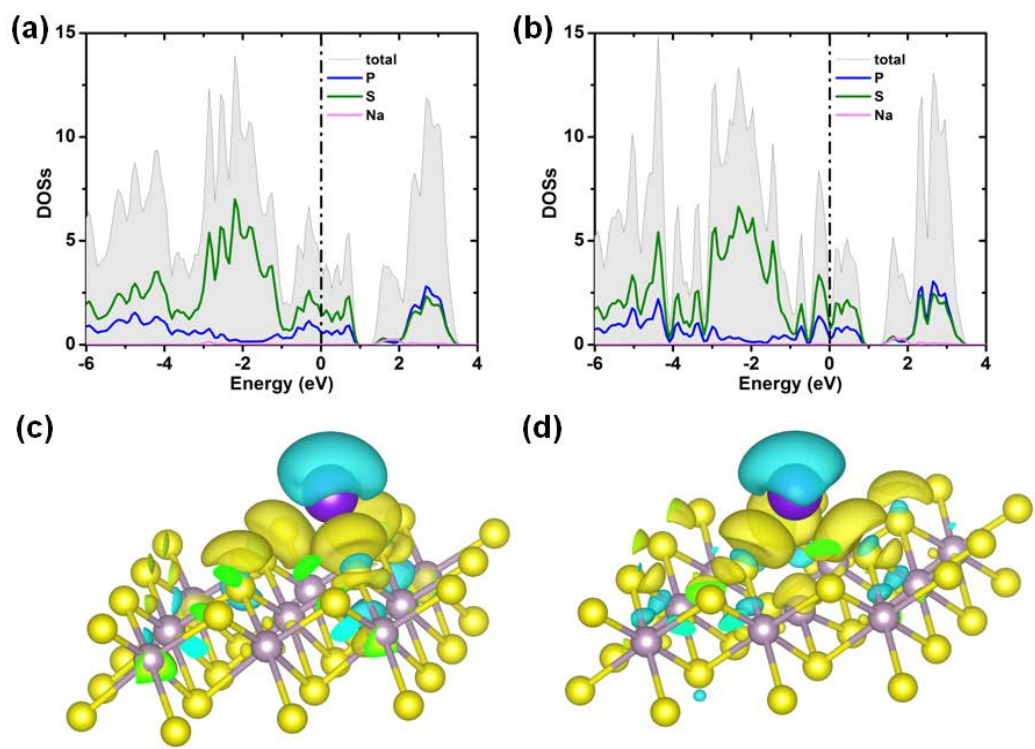

Figure 4 The density of states of NaPS$_2$ with Na atom at (a) V-site and (b) H-site. The charge density difference with the adsorption of Na atom at (c) V-site and (d) H-site. The gold color (i.e., 0.005 eÅ$^{-3}$) in the plot indicates an electron density increase and the cyan color (i.e., 0.005 eÅ$^{-3}$) indicates an loss.

The Na ion diffusion energy barrier has a great effect on the charging and circuit rate capacity of SIBs. According to the above results, the data of adsorption energy shows a small difference of a Na ion at V- and H-sites (0.09 eV), which is due to existence of a similar S environment at their adsorption sites. So the Na ions are expected to have a minimum energy path for Na diffusion from a V-site to the nearest neighboring V-site via the H-site (path I). Furthermore, the path from V-site to adjacent V-site directly is taken into consideration (path II). The paths and the relative energy profile are shown in Figure 5, where the adsorption energy of Na atom at the V-site is taken as reference. The calculated diffusion barrier of Na ion along the path I is 0.17 eV, which is the lowest of two possible circumstances. It can be explained that this path can reduces the influence of the energy variation at different sites.

†Correspondence should be addressed to: Chunying Pu (puchunying@126.com.)

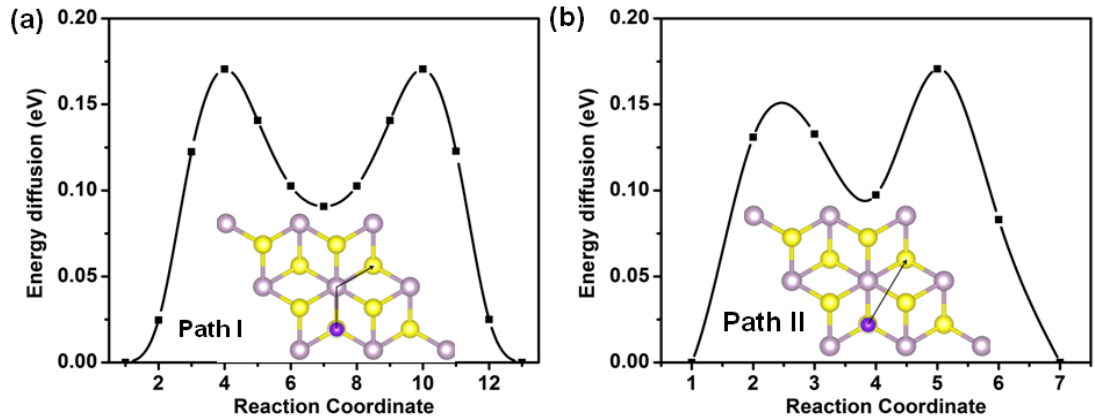

Figure 5 Relative energy profile for the diffusion of Na ion on the surface of monolayer PS$_2$ along path I (a) and path II (b). The inset is the top views of the trajectory of Na ion diffusion over the surface of PS$_2$ monolayer.

The important parameters of NIBs as electrode materials are the open circuit voltage (OVC) and theoretical storage capacity. The theoretical storage capacity is directly concerned with the number of adsorbed atoms. The intrinsic advantage of the monolayer materials is double Na storage capacity through adsorbing multilayer Na on both sides. The average adsorption energy layer by layer can be obtained by: $E_a = (E_{Na_{8x}P_4S_8} + Na_{8(x-1)}P_4S_8 - 8E_{Na})/8$. Here x represents the number of layers of Na atom. The negative adsorption energy means the adsorption of x layers is accessible and the maximum storage capacity can be obtained by $C_M = \frac{mF}{M_{PS_2}}(mAhg^{-1})$ where m is the number of adsorbed Na ions on the PS$_2$ per formula unit, F (26801 mAhg$^{-1}$) is the Faraday constant, and $M_{PS_2}$ is the molar mass of PS$_2$ per formula unit. As shown in Figure 6a, three layers of Na atom on each sides for Na ions adsorption on 2×2×1 supercell. The first Na atom layer is located at the V-site and the average adsorption energy is -0.45 eV. For the second layer, Na atoms prefer to be adsorbed at the T-site and the average adsorption energy becomes -0.06 eV. As for the third layer, Na atoms positions are the same as the first layer and the average adsorption energy is -0.02 eV. The adsorption of three-layer Na atoms on each side of $PS_2$ monolayer can be understand by the distribution of the dispersive electron cloud (Figure 5b) acting as S ions. Electron clouds distributed around Na ion can effectively alleviated repulsive interactions between Na ions. The electron cloud and the negative adsorption energy indicate that at least two layers of Na atoms can be adorbed on the $PS_2$ monolayer, which make the $PS_2$ monolayer a high capacity. The maximal theoretical capacity of the $PS_2$ monolayer can reach 1692 $mAhg^{-1}$, which is only lower than the capacity for borophene (1984 $mAhg^{-1}$),[31] but is even larger than other reported 2D disulfides (e.g.146 mAh/g for $MoS_2$,[62] 466 and 233 mAh/g for $VS_2$,[63,64] 479 mAh/g for $TiS_2$[53]) and comparable to that of $NiC_3$(~1698 $mAhg^{-1}$).[65] During the Na ions intercalation process, the lattice constants in the x-y plane only experience a tensile strain about 12.7%, which can comparable to the typical values that below 10% are acceptable.

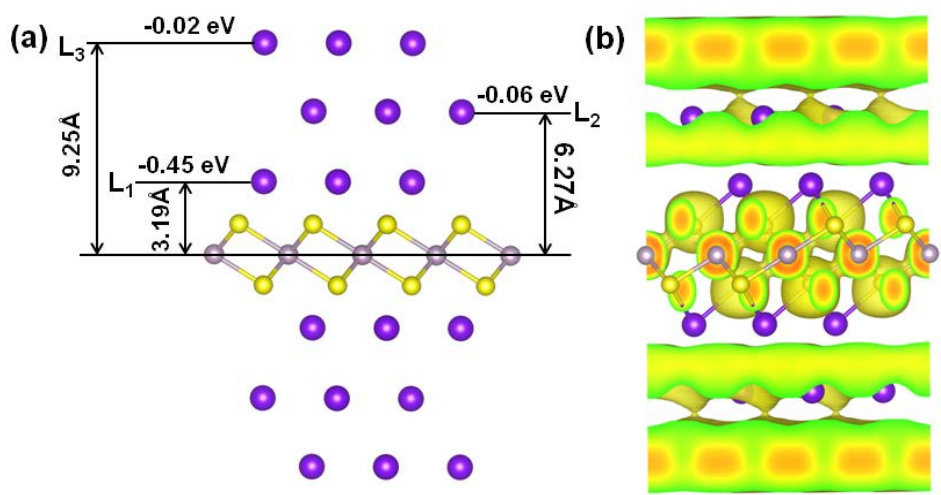

Figure 6 (a)The side view of the atomic structure of Na-intercalated $PS_2$ monolayer, where three Na layers adsorbed on each side monolayer. (b) The electron local

†Correspondence should be addressed to: Chunying Pu (puchunying@126.com.)

function (ELF) of PS$_2$ monolayer with three-layers of Na atom adsorbed on each side.

In addition, open circuit voltage was also computed to estimate the performance of the PS$_2$ monolayer as an anode material. A low OVC of the anode imply the possibility of a high net cell voltage. The charge/discharge process of PS$_2$ monolayer can be described as $PS_2 + nNa^+ + ne^- \leftrightarrow Na_nPS_2$. For this reaction, when the change of volume and entropy during the adsorption process are neglected, the average open circuit voltage can be defined by $V_{ave} = (E_{PS_2} + nE_{Na} - E_{Na_nPS_2})/ne$, where $E_{PS_2}$ and $E_{Na_nPS_2}$ are the total energies of the PS$_2$ monolayer before and after the adsorption of Na atom, $E_{Na}$ is the energy per Na atom in its stable bulk structure of Na metal, n is number of adsorbed Na content on a 2×2×1 supercell of PS$_2$ monolayer. With the increase of the adsorbed Na concentration from 8 to 24 atoms on the 2×2×1 supercell, the OVC decreases from 0.45 to 0.18 V. The dropping voltage with the increasing Na ion concentration has also been reported for other anode materials.[11,66] In summary, PS$_2$ possesses excellent stability and superior qualities and ultrahigh capacity for application as anode material in SIBs.

## 4. Conclusion

In conclusion, we have proposed the two dimensional PS$_2$ monolayer with dynamic, thermodynamic, and mechanical stability by first-principles investigations. Its electronic properties are investigated, and it keeps metallic feature before and after the adsorption of Na atom and thus has good electric conductivity. The relatively small lattice changes during the intercalation of Na and three Na atoms adsorbed steadily in each layer on both sides of the PS$_2$ surface reveals good recyclability and the maximum storage capacity of the PS$_2$ monolayer can reach to 1692 mAhg$^{-1}$ for NIBs, which is quite high among two dimensional materials. The calculated diffusion energy barrier of 0.17 eV for Na indicates that $PS_2$ monolayer can possess fast charge/discharge rates for Na atom in the SIBs. Moreover, we find that the superconducting state is characterized by an electron-phonon coupling constant and a superconducting critical temperature of 10.0 K. Our calculated results shown that two dimensional $PS_2$ monolayer can be applied as nanoscale superconductor and a promising as electrode materials, and awaits experimental confirmation.

## Acknowledgements

This research was supported by the National Natural Science Foundation of China (grant no. 51501093); The Henan Joint Funds of the National Natural Science Foundation of China (grant nos U1904179 and U1404608); The Key Science Fund of Educational Department of Henan Province of China (No. 20B140010).

†Correspondence should be addressed to: Chunying Pu (puchunying@126.com.)

†Correspondence should be addressed to: Chunying Pu (puchunying@126.com.)